# Software Engineering for IoT-Driven Data Analytics Applications

Aakash Ahmad[1], Mahdi Fahmideh[2], Ahmed B. Altamimi[1], Iyad Katib[3], Aiiad Albeshri[3]
Abdulrahman Alreshidi[1], Adwan Alanazi[1], Rashid Mehmood[4]

[1] College of Computer Science and Engineering, University of Ha'il, Saudi Arabia; a.abbasi@uoh.edu.sa, ab.altamimi@uoh.edu.sa, ab.alreshidi@uoh.edu.sa, a.alanazi@uoh.edu.sa

[2] School of Computing and Information Technology, University of Wollongong, Australia; mahdi@uow.edu.au

[3] Department of Computer Science, FCIT, King Abdul Aziz University, Jeddah, Saudi Arabia; iakatib@kau.edu.sa, aaalbeshri@kau.edu.sa

[4] High Performance Computing Center, King Abdulaziz University, Jeddah, Saudi Arabia; RMehmood@kau.edu.sa

**ABSTRACT** Internet of Things Driven Data Analytics (IoT-DA) has the potential to excel data-driven operationalisation of smart environments. However, limited research exists on how IoT-DA applications are designed, implemented, operationalised, and evolved in the context of software and system engineering life-cycle. This article empirically derives a framework that could be used to systematically investigate the role of software engineering (SE) processes and their underlying practices to engineer IoT-DA applications. First, using existing frameworks and taxonomies, we develop an evaluation framework to evaluate software processes, methods, and other artefacts of SE for IoT-DA. Secondly, we perform a systematic mapping study to qualitatively select 16 processes (from academic research and industrial solutions) of SE for IoT-DA. Thirdly, we apply our developed evaluation framework based on 17 distinct criterion (a.k.a. process activities) for fine-grained investigation of each of the 16 SE processes. Fourthly, we apply our proposed framework on a case study to demonstrate development of an IoT-DA healthcare application. Finally, we highlight key challenges, recommended practices, and the lessons learnt based on framework's support for process-centric software engineering of IoT-DA. The results of this research can facilitate researchers and practitioners to engineer emerging and next-generation of IoT-DA software applications.

## I. INTRODUCTION

IoT as a network of interconnected devices has emerged as a disruptive technology and an enabling platform that interconnects heterogeneous things such as humans, systems, services, and devices in a smart environment [1, 2, 3]. A recent survey by Gartner predicts that by the year 2020, the world will become home to 20 billion internet-connected things with data produced, consumed, and processed by IoTs will funnel through IoT-DA platforms [1]. A rapid proliferation of IoT systems is primarily due to portable devices that unify hardware (embedded sensors) software (applications that manipulate sensors) and network (protocols that connect sensors) to enable things that collect, process, and exchange contextualised data [2]. Typical example of such contextualised data includes crowd-sensed traffic congestions or environmental pollution that can be captured by embedded sensors of mobile devices, manipulated by mobile applications, and transmitted over wireless networks [3]. IoT systems in general and IoT-DA applications in particular act as the backbone for data-driven smart cities and infrastructure initiatives across the globe such as in United States [4], Europe, and Asia [5, 6]. Software or hardware novelties for IoTs are vital, however; true potential and business value of IoTs lies with IoT-DA that derives useful intelligence from data captured by sensors and things for strategic decision making. Due to an inherent complexity and heterogeneity of IoTs, one of the critical challenges is to develop processes, algorithms, tools, and frameworks etc. for sustained engineering and development of IoT-DA applications and platforms [7].

### A. SE for IoTs

Software engineering principle and practices support the design, development, evaluation, and maintenance phases of complex and large scale software-intensive systems effectively and efficiently [8]. In recent years, SE research and development have mainly focused on addressing challenges that relate to the development of IoTs in the context of smart systems and cities [3, 7]. A recently published roadmap for the adoption of emerging technologies streamlines engineering efforts – processes and frameworks with tool support – for sustained development and economic viability of systems such as IoTs, blockchain, and mobile-cloud [1, 3]. SE for data analytics unifies science and engineering of analysing data by means of algorithms, tools, and technologies that transform raw data into useful information using software applications [7]. The International Data Corporation forecasts that by the year 2022 the revenue generated by big data analytics will reach approximately 274.3 billion USD [9]. Despite the financial benefits and technical capabilities, data analytics technologies can be difficult to develop and time consuming to maintain due complexities of exponentially growing data and escalating costs of scaling these solutions to real problems [7]. SE for IoT-DA can support strategic capabilities of enterprises by exploiting software tools, algorithms, and applications that collect data from IoT-driven sensors and deliver intelligent analytics to stakeholders. For example, an application that logs availability and performance of interconnected robots in industrial automation can algorithmically mine those logs to discover patterns as reusable knowledge, best strategies, and key insights for optimising process automation [10]. In the context of engineering IoTs [11], IoT application implementation differs from mainstream mobile or web application development due to specific requirements of IoTs such as device heterogeneity, things' reactivity, and software services mapping with hardware sensors [10, 11].

From system development and operation perspective, an IoT-DA application is composed of interconnected things (i.e., context sensors) that are orchestrated by software services (i.e., data and logic) to provide intelligence based on information that is fused from sensors into centralised repositories [11, 12]. A complex blend of hardware and software artefacts poses challenges for engineering and development of dynamic IoT applications that require reusable knowledge and best practices of software and system development [10]. Engineering life-cycle to develop IoT systems and IoT-DA applications must take precedence over ad-hoc and once-off efforts that lead to increased efforts, decreased quality, and ultimately inferior product delivery [2, 3]. One of the recently published reports [13] by Gartner suggests that ' … *standardisation of engineering and development processes is the key to support sustained growth and adoption of IoT-driven applications. Standardised processes and practices enable organisations to follow systematic and process-centric approaches to engineer IoT systems effectively and efficiently…*'. Researchers and practitioners argue about the application of traditional software development life-cycles (SDLCs) such as agile and iterative approaches to develop IoT systems and applications [10]. However, some recent surveys indicate that IoT systems represent a complex combination of hardware and software components that need customised methods and tools for their development [10, 11]. In addition to the technical challenges, IoT systems such as smart transportation system involves multiple domains and a diverse set of stakeholders including public organisations, policy makers, and citizens with varying requirements [14].

*B. SE processes for IoT-DA applications*

Researchers and practitioners in the domain of software and system engineering are continuously deriving taxonomies and frameworks to classify, compare, and evaluate existing methods, technologies, and processes that enable or enhance engineering lifecycle [15, 16, 17, 18]. To date, there does not exist any work on evaluating SE processes, practices, methods, and tools etc. to engineer software applications for IoT-driven data analytics. This research is a pioneering effort to streamline the process and engineering lifecycle - based on academic research and industrial solutions - for software applications that derive key intelligence from data ingested from IoT sensors. We propose that by unifying (i) software engineering processes, (ii) internet of things development, and (iii) data analytics methods, an engineering lifecycle can be adopted that supports design, development, operationalisation, and evolution of emerging and next generation of IoT-DA applications. The objective of this research is to *empirically identify and systematically evaluate the existing processes and their underlying practices (based on academic research and industrial solutions) to develop IoT-driven data analytics applications.* Criteria-based evaluation of the SE processes streamlines development activities and recommended practices to develop IoT-DA.

*C. Method and overview of the proposed solution*

We have used evidence-based software engineering approach to develop a criteria-based framework[1] that streamlines and objectively evaluates SE processes for IoT-DA applications. As in Figure 1, the research phases include:

- *Phase 1:* Develop a framework to evaluate SE processes and their underlying practices that enable SE for IoT-DA applications.

- *Phase 2*: Apply systematic mapping study to qualitatively select and document 16 processes (08 each from academic research and industrial solutions) that support SE for IoT-DA applications.

- *Phase 3*: Evaluate the SE processes (Phase 2) using the framework (Phase 1) for criteria based evaluation of development processes, tools, and other artefacts of SE for IoT-DA applications.

- *Phase 4*: Support process-centric development of smart healthcare system as a case study to present guidelines, recommendations, and lessons learnt for engineering IoT-DA applications.

The proposed research complements one of the recent community-wide initiatives on exploiting engineering practices to develop IoT-driven systems [2]. Primary research contributions include:

- Establishing a process-centric (a.k.a. reference model) view based on existing SE research and practices to support the engineering life-cycle for IoT-DA applications.
- Empirically developed framework for criteria-based evaluation of SE processes and their underlying activities for IoT-DA applications.
- Creating a repository of SE processes as a structured catalogue of solutions and development activities. Case study along with recommended practices and lessons learnt can help:

    - Researchers to understand state-of-the-art, analyse its strengths and limitations, and derive new theories or hypotheses to be tested in the context of SE for IoT-DA.

    - Practitioners to understand the extent to which academic research and its outputs (architectures, algorithms, patterns, prototypes etc.) can be leveraged to develop practical and scalable solutions for IoT-DA.

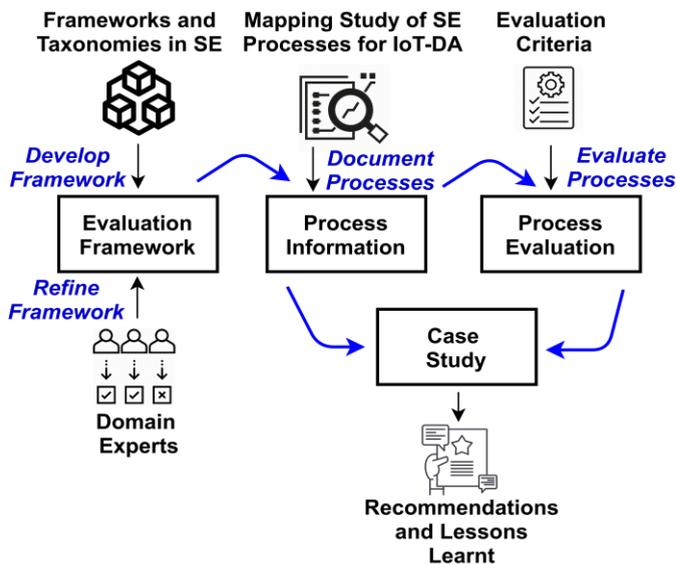

**FIGURE 1.** Overview of Research Method and Proposed Approach

**Organisation of paper:** Section 2 presents related research and reference model for IoT-DAs. Section 3 presents research methodology and evaluation framework. Section 4 presents the results of process evaluation. Section 5 presents a case study on IoT-DA application. Section 6 presents recommendations, lessons learnt, and threats to research validity. Section 7 concludes the paper.

## II. Related Research and Reference Model of SE for IoT-DA

Smart cities and societies provide state-of-the-art approaches for urbanisation built on smart applications, systems, and other infrastructures [19]. Internet of Things allow micro-level sensing of smart environments [20]. Millions of physical and virtual sensors generate invaluable data, i.e., big data [21], which is processed through various computational intelligence methods (e.g., as in [22]) to provide analytics capabilities for design, analysis and operationalisation of these smart environments [23]. However, we will see in this section that limited research exists on how IoT-DA applications are designed, implemented, operationalised, and evolved in the context of software and system engineering life-cycle. In this section, first we review related research and development in the context of SE for IoT-DA applications that helps us to justify the scope and contribution(s) of proposed work (Section 2.1). Review of the existing research and development also enable us to derive a reference model that conceptualises software engineering process for IoT-driven data analytics (Section 2.2). The terminologies and concepts introduced in this section are used throughout the paper based on illustrations in Figure 2.

### A. Related Research on SE for IoT-DA

We now present related research and development in the context of SE for IoT-DA. The discussion about related work is guided by the details in Figure 2. Figure 2 synergises already established principle and practices from different research domains to pinpoint the needs for SE methods and techniques that can be applied to data analytics systems that gather data from IoTs sensors. In the following, we discuss state-of-the-art research on evaluation frameworks in SE (Section 2.2.1), SE for IoTs (in Section 2.2.2) and IoT driven data analytics (in Section 2.2.3). The discussion about existing work helps us to identify the existing solutions and their limitations to justify the scope and contributions of proposed research in Table1.

### A.1 Evaluation Frameworks in SE

In software engineering, a number of frameworks have been used for criteria-based evaluation of software processes and their underlying activities such as software architecture, software evolution, and software evaluation. For example, the framework in [15] classifies and compares description languages for architecture-centric development of software-intensive systems. Other notable examples include classification and comparison driven evaluation of software evolution [16, 17] and evaluation [18]. The evaluation framework in [24] appraises existing cloud migration processes to objectively assess their similarities, weaknesses, and strengths. The framework and results of evaluation enable both the researchers and practitioners of cloud computing to assess the suitability of processes for the cloudification of legacy software. Existing frameworks are limited to evaluation of an individual activity of the SE process that lack (i) evaluation of the process lifecycle, (ii) application of SE process to emerging domains such as big data and IoTs. In recent years, these frameworks have also been used to evaluate the systems such as cloud [25] and IoTs [26]. The framework in [26, 27] can be seen as an initiative to evaluate SE process and lifecycle activities for IoT-driven systems. Specifically, the analytical framework in [27] evaluates and provides comparative analysis of nine well-known IoT architectures. The results of architectural evaluation provide insights to city leaders, architects, and developers to select the most appropriate architecture for implementing smart city systems. We summarise comparative analysis of existing evaluation frameworks and taxonomies (detailed above) with the proposed framework in Table 1 at the end of this section. The proposed framework extends the concepts and classification scheme from [26] to specifically evaluate SE processes and practices to engineer systems of IoT data analytics.

## A.2 Software Processes and Engineering for IoTs

In an IoT context, the concept of data fusion, i.e., things and devices producing, consuming, and exchanging large sets of data (volume, velocity, and variety of it) give rise to big data [21, 22]. IoT supported big data is the key enabler for data-driven intelligence that provide foundations for sustainable computing environments in the form of intelligent systems and smart cities [3]. In recent years, a number of research studies have focused on applying SE processes (principle and practices) focused on architectures, patterns, frameworks, and tool support to engineer IoT-based software systems [2, 3, 28]. A recently published road map to software engineering in IoT era suggests that SE research and development have mainly focused on addressing issues that relate to the development of IoTs in the context of smart systems and cities [11]. However, SE for IoT approaches primarily focus on modelling and implementation efforts only that overlooks process-centric approach and engineering life-cycle of IoTs systems [7]. The key to successful application of SE approaches to engineer IoT systems lies with abstractions on which IoT system and application engineering could be developed [7, 8]. These abstractions can encapsulate the analysis, design, and development phases of IoTs to develop architecture, frameworks and tools to engineer IoT systems and software. An example is ThingML approach that is inspired by UML based system modelling to addresses the challenges of distribution and heterogeneity in the Internet of Things [12]. ThingML supports model-driven engineering approach to develop high-level design models that drive coding, testing and evolution of IoT-based e-health solution. ThingML [12] as IoT specific modelling notation extends the traditional architecture description languages [15] for domain specific modelling of IoTs [12].

## A.3 Engineering of IoT-driven Data Analytics Applications

IoT devices and things represent the backbone for smart city systems - providing an infrastructure for digital services - in the form of smart transportation [29], healthcare [30], urban planning [14], and industrial automation [31]. Academic research and industrial solutions are in a continuous pursuit to empirically derive and validate processes, methods, and tools of software engineering, architectural patterns, design patterns and development methods to develop solutions for IoT-DA [3]. Some of the key industrial players for IoT infrastructures and solutions (e.g.; Google, Microsoft, Amazon) have developed solutions to support IoT-driven process automation, smart systems, and big data analytics [24, 27, 31]. The backbone for these smart systems is interconnected sensors that continuously produce and consume mission critical data to support the operations of the system [2]. Mission critical data refers to data and key information that supports the mission of a particular system. For example, in smart transportation system, data such as route planning and traffic congestion aims to support mission of smart transportation based on efficient route planning and execution [29]. Existing research is mainly focused on architectural abstractions [2], design patterns [12], and deployment models for IoT-DA. There is more focus on implementation but less focus on engineering cycle of the IoT-DA solutions. A process-centric approach enables an incremental development of the IoT-DA systems. The existing research also highlights that there is a need for tool-support to enable process automation, human roles and decision support to customise and supervise the development process [7, 30, 14].

***Conclusive Summary of Related Research:*** Table 1 provides comparison summary of existing frameworks and taxonomies to evaluate processes and artefacts of software engineering. The list of evaluation frameworks in Table 1 is not exhaustive but most relevant in terms evaluating SE specific processes and artefacts. Table 1 compares the state-of-the-art of evaluation frameworks based on (i) artefacts of evaluation, (ii) number of solutions evaluated, (iii) type of system under investigation and (iv) focus of evaluation. Each evaluation also shows the year in which the evaluation framework was put forward along with a reference to the framework. For example, [27] published in 2018 provides a framework for classification and comparison of architectural modelling phase by evaluating a total of 09 architectural models for IoT-driven smart city systems.

| Reference of Study | Artefact of Evaluation | Number of Solutions Evaluated | Type of System | Focus of Evaluation | Year of Publication |
|---|---|---|---|---|---|
| [15] | Software Design and Architecture | 10 | Component-based | Architecture modelling notations and description languages | 2000 |
| [16] | | 60 | Component and Service-based | Architecture maintenance and evolution methods | 2013 |
| [27] | | 09 | IoTs and Smart City | Smart city architectures | 2018 |
| [18] | Software Evaluation | 08 | Component-based | Architecture evaluation methods | 2004 |
| [25] | Software Evolution | 23 | Service and Cloud | Legacy to cloud migration processes | 2013 |
| [17] | | 32 | Component-based | Reuse knowledge for software maintenance and evolution | 2014 |
| [24] | | 43 | Cloud Computing | Legacy to cloud migration processes | 2016 |
| [26] | Software Process | 63 | IoTs and Smart Cities | IoT development platforms | 2018 |
| Proposed Framework | | 16 | IoTs and Data Analytics | IoT-driven Data Analytics | NA |

We can conclude that state-of-the-art on engineering and development of IoT software mainly focuses on application of existing process models to analyse, design, implement, and deploy IoTs for smart systems [10]. A recent survey on IoT development highlight that IoT driven applications represent a unique class of systems with blended hardware components and software services [26]. In IoT-DA applications, the challenges such as hardware software mapping, things analysis, sensor deployment, system elasticity require tailored processes and their underlying practices. The proposed research aims to empirically identify and systematically evaluate the existing processes and their underlying practices to develop IoT-driven data analytics applications that lacks in existing work. Criteria-based evaluation of the SE processes can help researchers and practitioners to understand the key challenges, recommended practices, methods, tools, algorithms, needs for process automation and human decision to develop emerging and new generation of IoT-DA applications.

### B. Reference Model for Engineering of IoT-DA Applications

Based on the review of existing research and development, summarised in Table 1, we have derived a reference model as illustrated in Figure 2. The reference model in Figure 2 acts as a system-level blueprint to unify software engineering process, IoT system, and steps of data analytics to conceptualise SE for IoT-DA. The reference model in Figure 2 can be interpreted by unifying (i) software engineering perspective and (i) IoT data analytics perspective, both detailed below.

### B.1 Software Engineering Perspective

SE for IoT-DA aims to apply engineering principle and practices to design, develop, operationalise, and maintain data analytic systems, software, and applications that ingest data from IoT devices and sensors [2, 7]. Specifically, real-time data consumed and produced by IoT devices such as traffic flows, health monitoring, and environmental conditions can be analysed to gather key intelligence for operationalising smart cities, systems, and infrastructures [3, 7]. As in Figure 2, software engineering process and its underlying activities lay the foundation for analysis, design, development, testing, deployment, and evolution (a.k.a. SDLC) of software intensive applications and systems [8]. The generic process in Figure 2 is adopted from [8] and customised to represent the de facto activities for SE (both from academic and industrial perspective) that support sequenced actions for an incremental engineering with possible iterations for process refinements. Tool support and human roles are complementary part of the process in order to support automation, customisation, and human decision support to execute and supervise the process for engineering software-intensive systems.

For example, at the system design phase, design patterns as reusable knowledge and best practices can be exploited to address the recurring challenges of developing IoTs. Specifically, the distributed architecture pattern can be applied that utilises communication hub as a mediator between interconnected devices and sensors [3, 7], as in Figure 2. Moreover, the layered deployment model helps with modularising data analytics application based on deployed sensors and devices (as front-end layer) to collect data that is processed and stored by cloud-based servers (as backend layer) [7].

### B.2 IoT-driven Data Analytics Perspective

A structured view of the IoT system and steps of data analytics show that raw data from IoT sensors can be collected and processed in an incremental manner to derive data-driven intelligence from IoTs [9]. In Figure 2, building blocks of IoT system are things such as a vehicle, digitise home, and service robot that are equipped with sensors and control software to coordinate with each other [7]. For example, in smart home settings the vehicle can coordinate the arrival time of resident to signal the service robot to finish up with cleaning and start brewing up evening coffee. Also, the vehicle can communicate with electronic appliances for maintaining appropriate lighting and room temperature before resident's arrival. The data consumed and produced by thing's sensors can be analysed to gather key insights such as traffic congestion for efficient routing, efficient home servicing (task priorities) by robot, and power usage patterns by home appliances for energy efficiency.

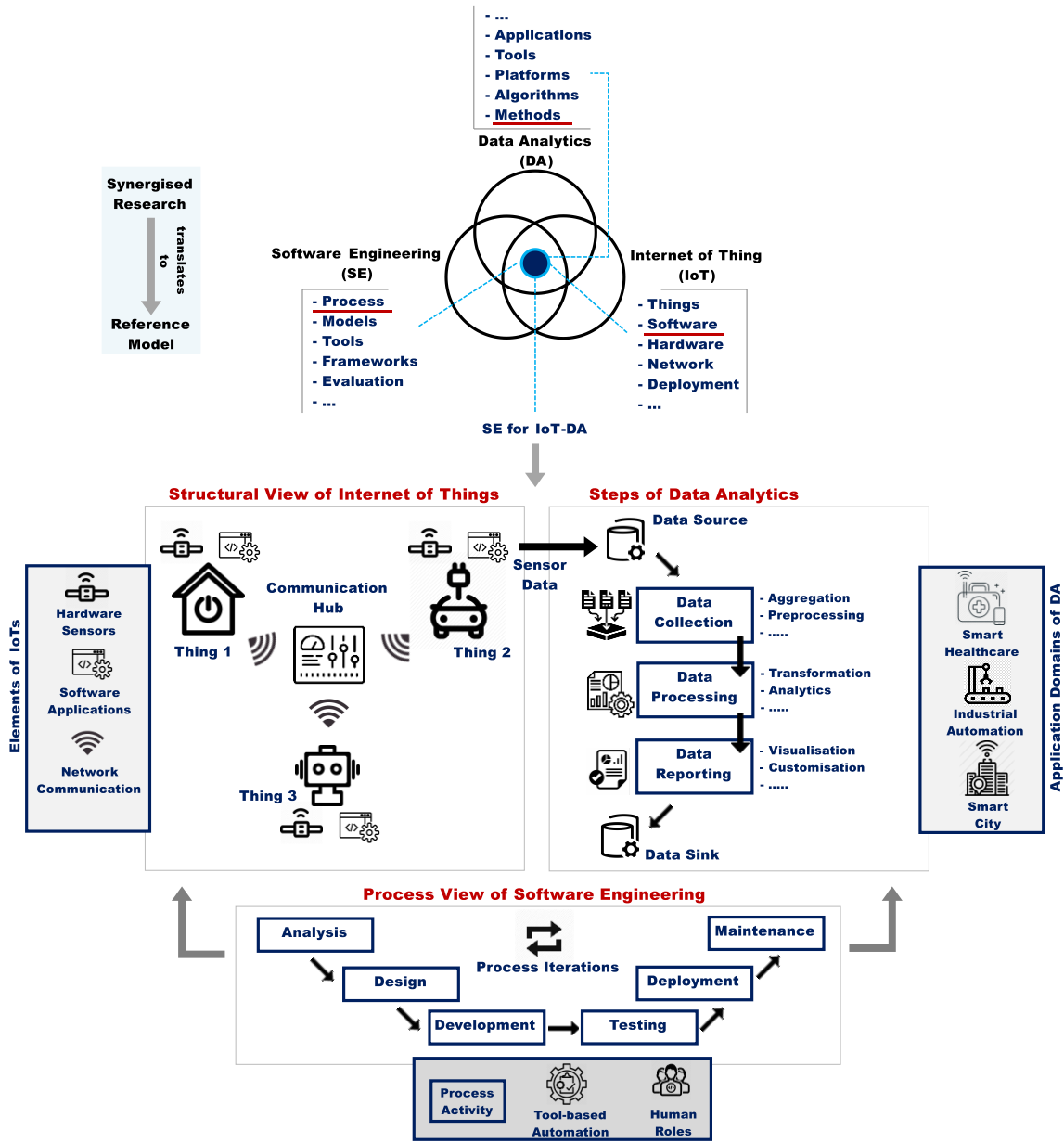

FIGURE 2. Reference Model of Software Engineering for IoT-driven Data Analytic

### B.3 Software Engineering Process to Develop IoT-DA Applications

Figure 2 highlight that developing and operationalising such a system poses challenges at multiple levels that include but not limited to the selection of deployment model, implementation algorithms, data security and privacy along with systems' accuracy, efficiency, and availability. By following a process-centric approach, software engineering methods can be applied in terms of architectural models, pattern templates, and algorithmic specifications to optimise resource management, computation demands, and operational efficiency of IoT systems. In the context of SDLCs [8, 24], SE process provides a blue-print of underlying activities and practices, supporting engineering life-cycle of software-intensive systems, that includes analysis, modelling, implementation, testing, and deployment of software applications under consideration. In this context, process as higher-level abstraction outlines: *what needs to be done?*, while process activities or practices demonstrate: *how that is to be done?*. For example, *software modelling* as a sub-process of the SE process contains an activity named *requirements specifications* that focuses on representing high-level design and specifications of the systems (in the context of functional requirements and quality attributes that must be satisfied) in the implemented software.

A typical argument is that instead of any (re-) inventions of development processes and engineering lifecycles for IoT software, traditional development cycles (e.g., agile, iterative SDLCs) must be used to develop IoT systems and applications [10]. However, recent surveys and empirical studies suggest that development of IoT-intensive software differ from traditional development practices for mobile or web application [11, 10]. Industrial solutions [13] also suggest that IoT systems exhibit specific requirements that need tailored process(es) with explicit support for device heterogeneity, things' reactivity, fault tolerance, system elasticity, multi-tenancy, and algorithmic manipulation of sensors [10, 11]. An example of SE for IoT-DA is presented in [29] that demonstrate how application of a layered software architecture and deployment pattern can support mobile cloud systems for crowd-sensing of traffic data. The example also demonstrates that well established principle and practices of SE (i.e., architectural modelling, requirements, and patterns) can support engineering of software applications for IoT-DA with required functionality and desired quality. Based on the smart transportation scenario in [29], some of the typical aspects of SE process and their impact on engineering IoT-DA applications include:

- **Software architecting** (*system blueprint*) as design level activity of SE process helps to structure the system into two layers namely mobile layer (front-end crowd sensing) and cloud layer (backend predictive analytics). In this scenario, the layered architecture model supports separation of functional concerns. Specifically, the front-end layer consists of portable and context-sensitive mobile devices that capture contextual information for traffic conditions such as date/time, location, traffic bottleneck on a given route. The backend layer supports storage and computation of traffic data that is being offloaded by mobile devices to cloud servers.
- **Architecturally significant requirements** (*quality attributes*) can pinpoint trade-offs, such as portability and context-sensitivity of mobile devices introduces some limitations in terms of limited processing, storage, and battery resources to perform any computation intensive tasks. This means that a backend system must be integrated as an additional architectural layer to which mobile devices can off-load crowd-sensed data for analytics and storage. Furthermore, quality attributes such as system's accuracy, efficiency, and availability must be evaluated to objectively assess the usefulness and applicability of the system.

- **Distribution pattern** (*reuse knowledge*) supports system deployment based on mobile cloud computing with event-based notifications. The distribution pattern deploys and operationalises crowd-sensing transportation system by means of a backend cloud server to perform analytics, communicates the results back to mobile devices, and ensures security and privacy of mobile data on cloud-based server.

## III. Research Method and Evaluation Framework
In this section, we overview the research methodology and evaluation framework. First, we discuss the research method in Section 3.1. We then introduce the framework and its building blocks for criteria-based objective assessment of the SE processes for IoT-DA applications in Section 3.2.

### A. Research Method to Develop and Refine the Evaluation Framework
We followed a four step research method with details of each individual step provided below as illustrated in Figure 3. The research method is primarily based on the guidelines to conduct systematic mapping studies and thematic analysis of topic under investigation [33]. Moreover, some existing frameworks and taxonomies such as [15, 16, 17, 18, 25, 26, 27] that classify, compare, and evaluate various phases of software engineering process and features of software-intensive systems are being used to derive the framework. Figure 3 is used for illustrative purposes to discuss different steps of the research methodology.

### A.1 Step 1 - Developing the Evaluation Framework
Figure 3 on the left illustrates Step A to develop the evaluation framework. Step A includes (1) investigating existing evaluation frameworks and taxonomies of SE, and (2) analysing the reference models and architectures for IoT systems. After conceptualising the framework, we engaged 7 domain experts (i.e., software engineers, software developers, project managers) with expertise in SE, IoTs, and data analytics. The feedback from domain experts helped us to refine and finalise the framework.

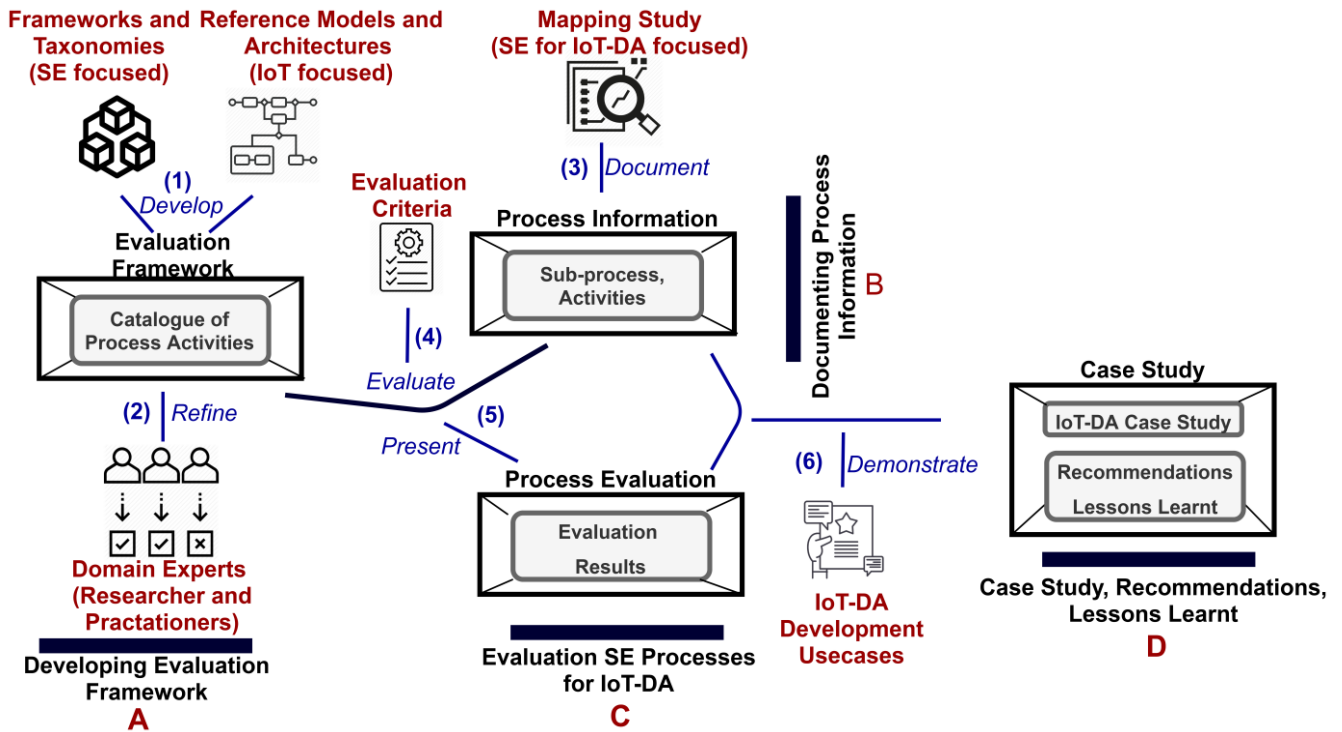

**FIGURE 3.** Illustrative Overview of the Different Steps of Research Method

Specifically, the framework was presented to the domain experts for their input and feedback followed by evaluation report and informal meeting that helped us to refine and finalise the framework.

### A.2 Step 2 - Documenting Process Information
Step B relates to documenting the process information as illustrated in the middle-top part of Figure 3. We conducted a mapping study to identify SE processes that have emerged from published academic research and documentation of industry specific solutions. Specifically, we had identified a total of 37 processes and qualitatively assessed them to shortlist 16 processes for evaluation. We selected 08 processes each from academic research and industry specific solutions. Based on reviewing existing taxonomies and frameworks in SE (e.g., [15, 16]), we derived 17 distinct criteria to objectively evaluate the process in terms of process planning, process execution, and process support for software engineering. Reviewing the most relevant processes both from academic research and industrial solutions helped us to create a process repository, as a central mechanism to store and retrieve process-centric information (see Table 2). The process repository accumulates all the process specific information (e.g., process activities, sub-activities) for engineering and development of IoT driven data analytic applications. The list of selected SE processes for IoT-DA applications is presented in Appendix A.

### A.3 Step 3 - Documenting the Evaluation Results
The next step in the methodology relates to documenting the results of evaluation that is illustrated in the middle-bottom part of Figure 3. The evaluation results highlight strengths and limitations of the processes and other useful information such as role of tool support and human supervision to automate and guide the process for IoT-DA.

### A.4 Step 4 - Case Study, Recommendations, and Lessons Learnt
The last step of the methodology focuses on presenting a case study on process-centric development of IoT-DAs, right side of Figure 3. Case study also helps to present some recommendations and lessons learnt from SE process and practices for developing IoT-DA applications.

### B. Framework for Evaluating Processes and Practices of SE for IoT-DA
We now discuss the framework in terms of its building blocks for the evaluation of the SE processes and practices for IoT-DA applications. Moreover, two distinct views of the framework help to distinguish between abstraction and instantiation of the framework detailed below. In the (software) engineering context, an evaluation framework is referred to as a hierarchical abstraction that organises and interrelates (i) artefacts being evaluated (e.g., tools, frameworks, (sub-) processes and their underlying activities) with (ii) criteria or features of evaluation[2].

---

[2]In the context of evaluation frameworks, the terms features and activities are virtually synonymous and often used interchangeably. We use the term activity as an individual criterion/step to support the process, whereas the term feature is used as a criterion to reflect the capability of a product.

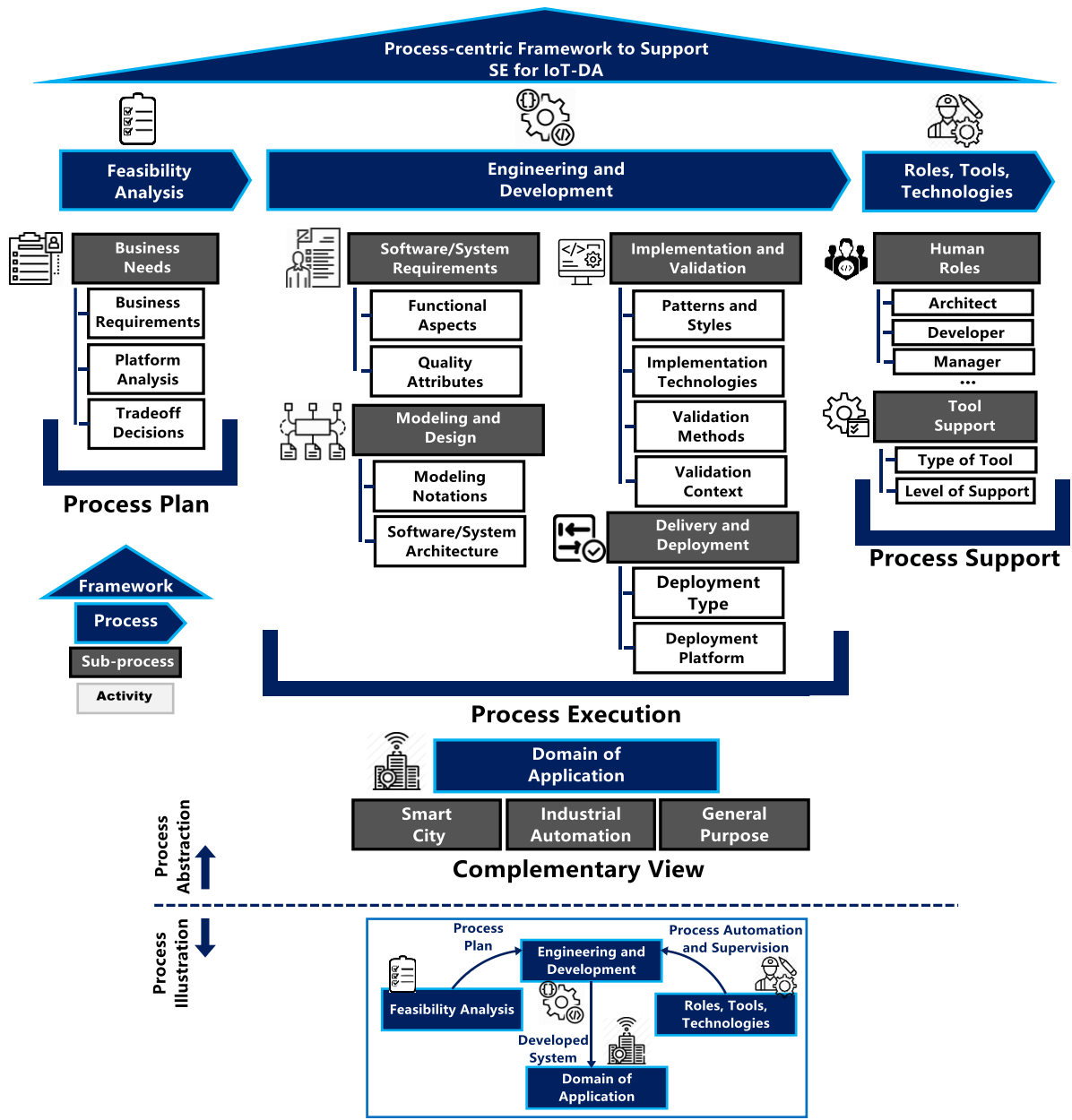

**FIGURE 4.** Overview of the Criteria-based Evaluation Framework
(Process Activities and Process Illustrations)

**B.1 Framework Views: Abstraction vs Instantiation**

The abstract view organises the processes, sub-processes and their underlying activities in a hierarchy. Figure 4 shows two distinct views of the framework in terms of framework abstraction and framework instantiation. The abstract view provides a blueprint of the framework that decomposes the core processes of SE research and practices in terms of (i) *Process Planning*: Feasibility Analysis, (ii) *Process Execution*: Engineering and Development, (iii) *Process Support*: Roles, Tools, and Technologies along with (iv) *Domain of Application*. The framework in Figure 4 is based on generic process for SE as in [8, 7, 15].

However, when the data in Table 2 (catalogue of process activities) is combined with framework views in Figure 4, sub-processes and activities can be interpreted specific to IoT-driven systems. For example, the SE process named Engineering and Development supports the sub-process of Modelling and Design. However, the activity Modelling Notation shows that in order to model IoT systems, specific modelling notations such as ThingsML [12] are required that can support features such as (i) abstraction of hardware and software mapping, (ii) things interaction, and (iii) specification of non-functional properties for IoTs.

Similarly, during the sub-process Delivery and Deployment, the Deployment Type activity can support mobile cloud distribution pattern where front-end mobile nodes (interconnected sensors) can offload context-sensitive data to backend cloud nodes (processing and storage servers). Framework instantiation shows how different processes and sub-processes interact to execute the SE process. For example, Figure 4 illustrates that feasibility analysis helps to create the process plan that supports engineering and development of the IoT-DA application.

## B.2 Representing and Mapping SE Processes and Practices in the Evaluation Framework

The framework for evaluating software engineering processes and activities to develop IoT-DA systems is presented in Figure 4. Referring to Figure 4, in software engineering process, modelling and design is a sub-process that can be evaluated based on existence of software/system architecture that acts as a blueprint for system wide design and implementation. In software engineering context, processes and activities are complementary to each other, i.e., process outlines step(s) to highlight engineering objectives (*what to do?*), while practices implements process(es) to achieve the engineering objectives (*how to do?*). For example, the sub-process software and system requirements elicitation supports the practices such as methods, tools, and technologies for identifying functional and quality attributes to streamline required functionality and desired quality of the software. The evaluation framework leans heavily towards classifying, analysing, and interpreting processes, their sub-processes, and underlying activities that enable engineering of data analytic systems and applications ingesting data from IoTs. Figure 4 presents a fine-grained hierarchical composition of the framework in terms of software engineering processes and their sub-activities to be evaluated. Table 2 complements Figure 4 to detail the criteria for evaluation in terms of a comprehensive catalogue of process activities. Table 2 acts as a structured catalogue to represent the evaluation criteria, query to retrieve process-centric information and relevant example. In addition, Table 2 also shows activities in the process that depend on each other. For example, in Table 2 the criteria Modelling Notation and Languages (MD1) aims to investigate: *Are there any modelling notations and/or languages being used to create system design?* Typical example of such modelling could be UML based notation and profiles to specify system design and architecture [12, 32]. The criteria MD1 depends on activity of collecting and specifying Functional Aspects (SR1) that needs to be modelled before system implementation (i.e., SR1 a prerequisite for MD1) in SDLCs.

## IV. Matrix and Results of Process Evaluation

We now present the artefacts and results of evaluation for SE processes and their underlying activities to develop IoT-DA applications, based on process matrix in Table 2. The process matrix in Table 2 is derived from evaluation framework (Figure 4) and process catalogue (Table 2) that comprises of four primary elements, referred to as the artefacts of evaluation. In order to support process evaluation:

- *Evaluation framework* in Figure 4 conceptualises different phases and activities of the SE process.

- *Process catalogue* in Table 2 complements the framework to provide fine-grained description of each process, its sub-process(es), and activities along with collaboration or dependencies between the activities.

- *Process matrix*[3] in Table 3 exploits the model and data from framework and catalogue to document process-centric information along horizontal (evaluation criteria) and vertical axis (process information). Matrix based approach enables structured representation of information for fine-grained evaluation of individual process.

### A. Matrix for Process Evaluation

The matrix for process evaluation is provided in Table 3 with artefacts of evaluation detailed below. Matrix-based evaluation driven by artefacts of evaluation reflects a collection of horizontal and vertical elements of the framework for an objective representation of process-centric data and fine-grained interpretation of process information, detailed below.

- **Sources of SE processes**. The processes, sub-processes, activities, and methods to engineer IoT-DA software emerge from two main sources namely Academic Research (📇) and Industrial Solutions (⚙). Academic Research represents well documented, peer-reviewed, published solutions that emerge as research and development initiatives by academia. In comparison, Industrial Solutions are commercially adopted operational systems and practices that are developed and adopted to address underlying business needs by enterprises. We reviewed a total of 16 processes and practices, 8 each from academia and industry. For example, in Table 3 CLOTHO [P3][4] represents academic research on IoT-driven monitoring and analytics of environmental data for smart disaster management. In comparison, SeeboIoT [P12] as an industrial solution support sensor data from industrial machines and processes for industrial automation and smart manufacturing.

---

[3]The term *Table 3* and *Process Matrix* can be used interchangeably both referring to same element of evaluation.
[4]The notation **[Pn]** (n is a number) represent references to processes that has been evaluated and presented in **Appendix A.** The notation also maintains a distinction between the referencing for bibliography and list of processes being evaluated.

**TABLE II** Catalogue of Process Activities (Assessing Processes, Activities, Sub-Processes)

| Process Activity | | Criteria and Query for Activity Evaluation | Activity Collaboration | |
|---|---|---|---|---|
| | | | Source | Dependent |
| **Feasibility Analysis (Process Planning)** | | | | |
| Business Needs and Trade-offs (BN) | 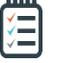 | **Criteria:** _Business Needs (BN1)_<br>**Query:** Are the business needs clearly specified to achieve system objectives?<br>**Example:** Cost-effective sensor-based monitoring of patient's health | × | × |
| | 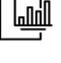 | **Criteria:** _Platform Analysis (BN2)_<br>**Query:** Are the platforms being used for system development and deployment objectively analysed?<br>**Example:** Deployment on distributed cloud model (Amazon Web Services (computations) and S3 Bucket (data storage)) | × | × |
| | 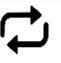 | **Criteria:** _Trade-offs Decisions (BN3)_<br>**Query:** Are the trade-offs (technical/non-technical constrains) being considered before system development?<br>**Example:** System's operational constraints (distribution vs performance vs availability etc.) | BN1, BN2 | BN3 |
| **Engineering and Development (Process Execution)** | | | | |
| System Requirements (SR) | 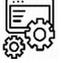 | **Criteria:** _Functional Aspects (SR1)_<br>**Query:** Are the requirements corresponding to system functionality being collected and analysed?<br>**Example:** System should have collected patient's health data using pulse sensors. | BN1 | SR1 |
| | 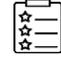 | **Criteria:** _Quality Attributes (SR2)_<br>**Query:** Are the requirements corresponding to system quality being analysed and prioritised?<br>**Example:** Security and privacy of patient's health critical data should be maintained. | BN2, BN3 | SR2 |
| System Modelling and Design | 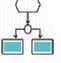 | **Criteria:** _Modelling Notations and Languages (MD1)_<br>**Query:** Are there any modelling notations and/or languages being used to create system design?<br>**Example:** Unified Modelling Language (UML) based notations and profiles for system design and architecture | SR1 | MD1 |
| | 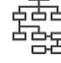 | **Criteria:** _Software/System Architecture (MD2)_<br>**Query:** Any software/system level architecture being created before implementations?<br>Example: Layered architecture for sensor-based health monitoring (i.e., data storage, sensor manipulation, and user interfacing layers) | SR1 SR2 | MD2 |
| Implementation and Validation (PV) | 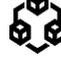 | **Criteria:** _Software Patterns and Styles (PV1)_<br>**Query:** Any software patterns or styles being applied to support reusability of code and quality of software?<br>**Example:** Publish-Subscribe patterns implemented by system to get notifications from health monitoring sensors. | SR1 SR2 | PV1 |
| | 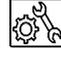 | **Criteria:** _Implementation Tools and Technologies (PV2)_<br>**Query:** What tools and technologies have been used to implement the system?<br>**Example:** AWS Developer Tools have been used to develop Amazon Web Services and Data Storage (S3 Bucket) | × | × |
| | 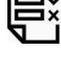 | **Criteria:** _Validation Methods (PV3)_<br>**Query:** What validation methods are employed to evaluate the system and validate its functionality and quality?<br>**Example:** Case study with user survey-based evaluation of the health monitoring system | SR1 SR2, MD3 | PV3 |
| | 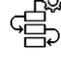 | **Criteria:** _Validation Context (PV4)_<br>**Query:** What is the context (environment) of system validation?<br>**Example:** Controlled experiments, simulations, and sample of real patients being used to validate the system. | SR1 SR2 | PV4 |
| Delivery and Deployment (DD) | 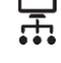 | **Criteria:** _Deployment Type (DD1)_<br>**Query:** What is the deployment model being used to deliver the system?<br>**Example:** Distributed deployment (computation and storage on cloud servers) with user interfacings (hand-held mobile devices). | BN2 SR2, MD2 | × |
| | 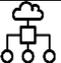 | **Criteria:** _Deployment Platform (DD2)_<br>**Query:** Which platform is being used to deploy the system?<br>**Example:** Web-service based deployment based on Amazon Web Services (AWS) platform | BN2 | DD2 |
| | | | | |
| Human Roles (HR) | 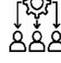 | **Criteria:** _Type of Role (HR1):_<br>**Query:** What type of human roles are involved to support the process?<br>**Example:** Management-oriented (Project Manager, Consultant), Technical (Software/System/Process Engineer)., Other (Testing User) | × | × |
| Tools for Engg and Dev. (TL) | 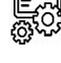 | **Criteria:** _Type of Tool Support (TL1):_<br>**Query:** What tools are being used and what is their source classification (Open/Closed source) for the tools being used?<br>**Example:** AWS Toolkit for Eclipse is used to develop health monitoring services. The Source type is Open Source | MD2, PV1 | TL1 |
| | 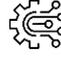 | **Criteria:** _Level of Tool Support (TL2):_<br>**Query:** What is the level of automation provided by the tool to develop the system?<br>**Example:** AWS Toolkit for Eclipse supports semi-automated development (manually created design can be auto-tested) | × | × |
| **Domain of Application (Complementary View)** | | | | |
| Domain of Application | 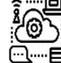 | **Criteria:** _Domain of Application (DA1)_<br>**Query: What is the domain of application for the developed system**<br>**Example:** Civil Services (Healthcare, Education), Industrial (Manufacturing, Assembling),General Purpose (Crowd-sensing) | BN1 SR1, MD2 | DA1 |

- **Process Lifecycle** can be divided into three main phases namely process planning (a.k.a. feasibility analysis), process execution (engineering and development), and process support (provided by human roles, tools, and technologies). Complementary view to process life cycle is domain of application that represents the area to which the solution can be applied. For example, the domain of application for CLOTHO [P3] is disaster management by analysing emergency scenarios and their impact in smart city context.

- **Evaluation Criteria** represents individual element to objectively evaluate the processes and practices. In Table 3, there are a total of 17 criterion to evaluate the process lifecycle that are distributed as *A. Feasibility Analysis* (03 criterion), *B. Engineering and Development* (10 criterion), *C. Roles, Tools, and Technologies* (03 criterion), and *D. Domain of Application* (01 criteria). For example, the criterion Quality Attributes (SR2) evaluates the engineering and development phase of IoT-DA system based on specification of quality attributes or non-functional requirements that must be satisfied by the system.

- **Levels of Attainment** reflects the strength of an individual activity and the process. There are a total of 5 levels including Optimal ( = 4.0), Satisfactory ( = 3.0), Partial ( = 2.0), Poor ( = 1.0), and None ( = 0). Each level is assigned with a numerical value ranging from 4 to 0 (from Optimal to None, respectively). The numerical values provide a quantified average of the evaluation score. The quantification can be an individual activity of the process in all the evaluated solutions, or quantification can be of all the activities in a single solution. For example, the total quantified value (i.e., evaluation score) for the activity named Functional Aspects (SR1) for all the processes is 2.0 (visually denoted as ( )), as in Table 3. In comparison, the quantified value based on all activities for an individual process named Industrial Analytics (IIoT) [P4] is 1.0 ( ). As in Table 2, a framework named CityPulse [P1] has satisfactory representation of the functional aspects in terms of (a) efficient route planning, (b) vehicular coordination to develop a smart transportation system. However, there are poor trade-off analysis for solution in terms of performance and availability of the system.

### B. Interpretations of Results for Process Evaluation
The interpretation of the results for process evaluation can be divided into two main types namely structured and unstructured. The difference between two types, detailed below, is the granularity of information that can be achieved. Structured results are derived based on pre-defined queries from Table 3 such as *is there any system/software architecture being created for the solution before implementation(s)?* The query to the process aims to identify the existence of design and/or architecture model in system engineering and development phase to evaluate the criteria MD2. In the following we use examples to demonstrate how process matrix in Table 3 can be used to get structured information regarding SE processes and practices in the context of IoT-DA. The examples below demonstrate how to interpret the results based on specific examples of process solutions being evaluated.

- *Are the requirements corresponding to system functionality being collected and analysed (in ITS solution)?* The result to this query could be retrieved by locating the ITS [P7] solution that highlight that functional requirements such as analysing traffic specific data and data driven transportation management have been explicitly defined.

- *What tools and technologies are used to implement the system (Amazon IoT Analytics)?* Amazon IoT Analytics solution [P14] uses Amazon Platform such as Amazon Web Services for data analytics from IoT devices. The data to be processed is stored and managed by Amazon Simple Storage Service (S3), whereas the Structured Query Language (SQL) has been used for querying IoT devices to extract specific data.

In comparison to the structured analysis, un-structured results can be interpreted based on specific information that needs to be extracted. For example, anyone interested to know about the level of tool support can identify Google IoT [P15] and ThingsSpeak [P10] as existing solutions that exploit maximum tool support for the engineering and development of IoT-DA applications. The following is further exemplification of how the process matrix could be used to analyse unstructured results.

- *What engineering activities are most and least focused in the development life-cycle for IoT-DA applications?* To find an answer to this question, a high-level view of the process matrix indicates that during IoT-DA application lifecycle, engineering activities that are most focused are business needs and platform analysis as part of the feasibility analysis. During the engineering and development phase, activities like specification of functional requirements, software design and architecture, and validation methods are most focused activities. In comparison, there is much less focus on trade-off decisions (cost-risk analysis, technical/non-technical constraints) on IoT system being developed. Modelling notations are focused less and level of tool support is not much supported.

**TABLE III** Matrix for Evaluation of SE Processes for IoT-DA Applications

**Criteria Fulfilment**

- Optimal (4)
- Satisfactory (3)
- Partial (2)
- Poor (1)
- None (0)

**Source of the Process**
- Academic Research
- Industrial Solution

**Process Support**
- Human role in the process activity
- Tool support to execute the process activity

| Evaluation Criteria | CityPulse [P1] | IoST Framework [P2] | CLOTHO [P3] | IIoT [P4] | Rathore et al. [P5] | Analytical Everywhere [P6] | ITS [P7] | IoTMobair [P8] | TCS Sensor Data Analytics [P9] | ThingSpeak [P10] | Industrial Analytics (IIoT) [P11] | SeeboIoT [P12] | Fujitsu IoT Analytics [P13] | Amazon IoT Analytics [P14] | Google IoT [P15] | Azure Data Explorer [P16] | Evaluation Score (criteria) |
|---|---|---|---|---|---|---|---|---|---|---|---|---|---|---|---|---|---|
| **Feasibility Analysis (Process Planning)** | | | | | | | | | | | | | | | | | |
| BN1 | | | | | | | | | | | | | | | | | 2.5 |
| BN2 | | | | | | | | | | | | | | | | | 1.5 |
| BN3 | | | | | | | | | | | | | | | | | 0.5 |
| **Engineering and Development (Process Execution)** | | | | | | | | | | | | | | | | | |
| SR1 | | | | | | | | | | | | | | | | | 2.0 |
| SR2 | | | | | | | | | | | | | | | | | 1.5 |
| MD1 | | | | | | | | | | | | | | | | | 0.75 |
| MD2 | | | | | | | | | | | | | | | | | 1.5 |
| PV1 | | | | | | | | | | | | | | | | | 1.5 |
| PV2 | | | | | | | | | | | | | | | | | 2.5 |
| PV3 | | | | | | | | | | | | | | | | | 2.0 |
| PV4 | | | | | | | | | | | | | | | | | 1.75 |
| DD1 | | | | | | | | | | | | | | | | | 2.5 |
| DD2 | | | | | | | | | | | | | | | | | 1.5 |
| **Roles, Tools, and Technologies (Process Support)** | | | | | | | | | | | | | | | | | |
| HR1 | | | | | | | | | | | | | | | | | 0.75 |
| TL1 | | | | | | | | | | | | | | | | | 1.5 |
| TL2 | | | | | | | | | | | | | | | | | 0.75 |
| **Evaluation Score** | 2.5 | 1.75 | 1.5 | 1.0 | 1.5 | 2.5 | 1.5 | 1.5 | 1.0 | 1.25 | 1.0 | 1.5 | 0.75 | 2.0 | 2.25 | 1.5 | |
| **Domain of Application (Complementary View)** | | | | | | | | | | | | | | | | | |
| DA1 | Smart Transportation | Environmental Monitoring | Disaster Management | Smart Healthcare | Urban Planning | Smart Transit | Smart Transportation | Environmental Monitoring | Enterprise Analytics | Smart City | Industrial Automation | Industrial Manufacturing | Stream Analytics | General Purpose | Industrial Automation | Smart Transportation | |

- *What are the prominent tools and technologies for the engineering and development of IoT-DA applications?* Prominent tools and technologies refer to tool chains and enabling technologies that are being used more frequently to support process automation, as per findings of reviewing and evaluating the processes. Specifically, tool represents software artefacts that can support individual activities or sub-processes to develop IoT-DA applications. For example, the tool named Eclipse IoT is among the most frequently used tool for creating the design and implementing source code for IoT-DA systems. In comparison to tools, technologies represent an umbrella of platform and underlying tools that provide an infrastructure to develop, deploy, and run IoT applications using a particular technology. For example, Amazon Web Services (AWS) as a collection of on-demand cloud computing technologies to operationalise, deploy and host IoT-DA applications, managed via pay-per-use services. For example, the solutions [P14] used the AWS technologies to perform analytics based on data storage with Amazon Simple Storage Service (S3).

Based on the details in evaluation matrix in Table 3, the results can be interpreted in different ways and dimensions in the context of information required. This means that the evaluation matrix provides foundation to analyse a particular process or its underlying activities to objectively assess the strengths and limitations of process-centric approaches for engineering and development of IoT-DA applications. An exhaustive presentation of evaluation results is not possible, however; details and examples above facilitate researcher and practitioners to interpret the results as per their requirements. In general, the overall process support for IoT-DA applications is weak with lack of tools and human decision support. Based on the evaluation results, we present a case study and outline some recommendations that highlight the lessons learnt to develop emerging and futuristic solution for IoT-DA applications.

## V. Case Study and Lessons Learnt for Engineering IoT-DA Application

We now present the case study on process-centric engineering and development of an IoT-DA application. We also discuss the experiences and lessons learnt based on case study-based approach to application development. Figure 5 provides a visual illustration of the process and its decomposition into various activities. Table 4 complements Figure 5 to provide structured analysis of each process, its sub-process and underlying activities that support system development.

### A. Process-driven Engineering of Connected Smart Healthcare Application

We follow a case study based approach to demonstrate process-centric development of an IoT-DA application. In software engineering context, process-centric development refers to an incremental approach to support the modelling, design, development, and deployment of the system under consideration [8, 32]. The Smart and Connected Healthcare (Health-Connect) case study is part of smart city initiative that enables individuals to exploit on-body (portable and connected) sensors that frequently monitor health signals originating from human body. These signals related to body temperature, pulse rate, and blood pressure etc., are captured by sensors deployed across human body (on-body IoTs), and are communicated to health analytics server as in Figure 5. Health analytic server as cloud-base storage and processing infrastructure generates health analytics in terms of a Health Profile. Health Profile is medically critical document that contains documented details of an individual's health conditions such as heart issues, blood pressure, and dietary plans. In addition to the analytics, the health profile(s) stored at server could be shared with other medical professionals for consultation. The ultimate goal of Health-Connect system is to exploit sensor-based data to analyse, profile, and manage connected and smart health care. In the following, Health-Connect case study and its development process, decomposed into three sub-processes with seven underlying activities (from Figure 3), are being analysed in the context of SE process (cf. Figure 2). In the following illustrations from Figure 5 and structured information from Table 3 are used to summarise three main phases of the process in the context of the Health-Connect case study.

### B. Feasibility Analysis for Health-Connect System

Table 4 is used to capture and present all the process-centric information for the engineering and development of the Health-Connect case study. Table 2 structures information with an explicit record for 07 items including *Process*, *Sub-process*, *Activity* of the sub-process, *Outcome* of each activity, *Human Roles* involved in the activity, along with *Tools and Technologies* that support the activity. In addition, *Domain of Application* is also detailed. For example, contextualising the case study from Figure 5, in Table 4, the System Modelling and Design process has one of the sub-processes named software architecture with an underlying activity to create layered architecture model for Health-Connect system. The layered architecture model is composed of two distant layers namely Health Sensing Layer (front-end health monitoring with IoT sensors) and Health Analytics Layer (backend health data processing layer with cloud server). The outcome of the (sub-) process and its underlying activity is a layered architecture for the Health-Connect.

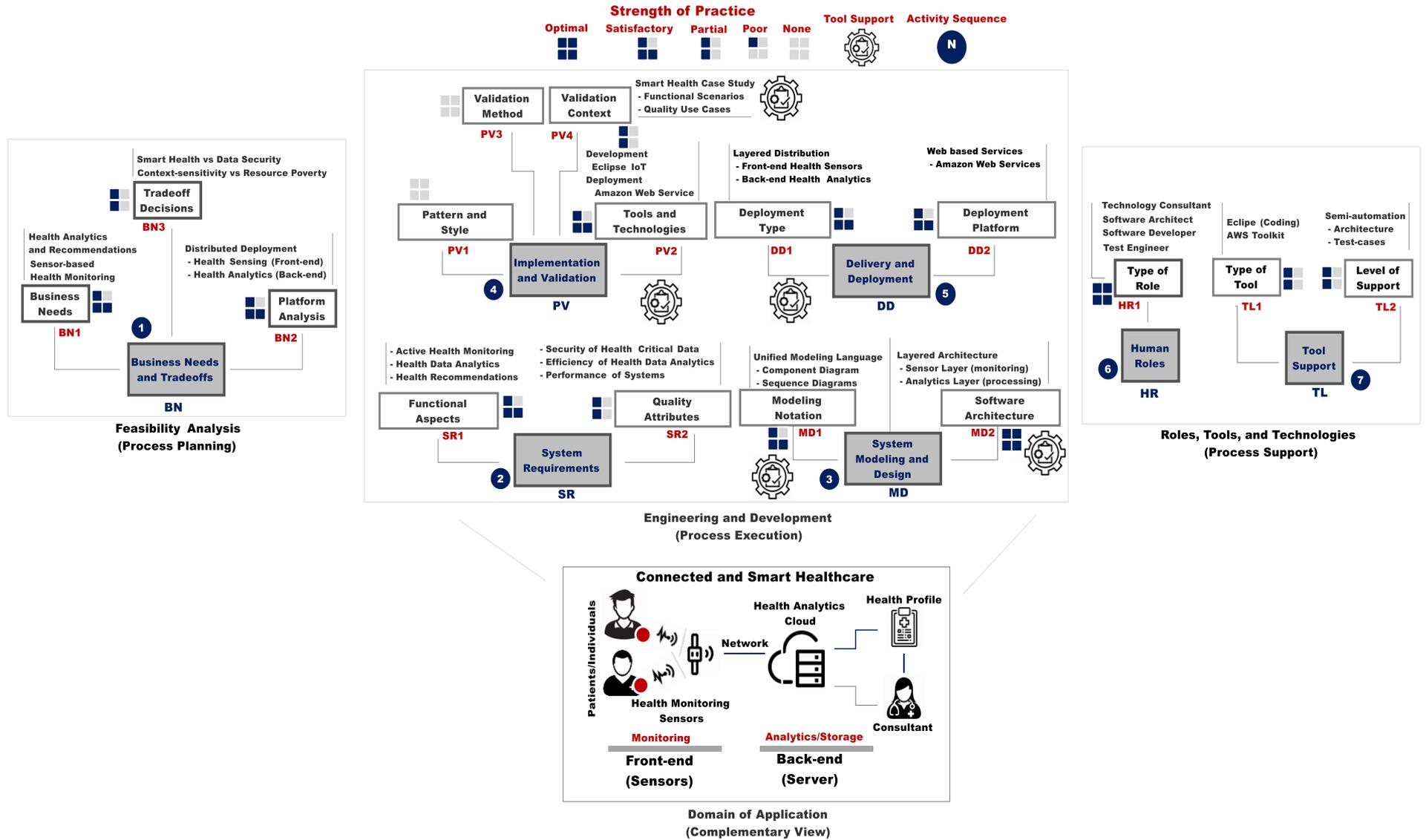

**FIGURE 5.** Overview of Process-centric Development of the HealthConnect Case Study

TABLE IV Overview of the Processes, Activities, and Specific Aspects of Health-Connect Case Study

| Process | Sub-process | Activity | Outcome | Human Role(s) | Tools and Technologies |
|---|---|---|---|---|---|
| **Feasibility Analysis** | | | | | |
| Business Needs and Trade-offs | Business Requirements | - Health Analytics and Recommendation<br>- Sensor-based Health Monitoring | None | Technology Consultant | None |
| | Platform Analysis | Distributed Deployment and Processing<br>- Health Sensing (Front-end)<br>- Health Analytics (Back-end) | Deployment Diagram | - Software Architect<br>- Test Engineer | Any software modelling tool (Microsoft Visio, Eclipse Modeling Framework, etc.) |
| | Trade-off Decisions | - Smart Health vs Data Security<br>- Context-sensitivity vs Resource Poverty | Cost/Benefits Report | Technology Consultant | None |
| **Engineering and Development** | | | | | |
| Software/System Requirements | Functional Aspects | - Active Health Monitoring<br>- Health Data Analytics<br>- Health Recommendations (Profiles) | Requirements Specification Document<br>- Functional Specifications<br>-Architecturally Significant Requirements | - Software Architect<br>- Test Engineer | None |
| | Quality Attributes | - Security of Health Critical Data<br>- Efficiency of Health Data Analytics<br>- Performance of Systems | | | |
| System Modelling and Design | Modelling Notation | Unified Modeling Language<br>- Component Diagrams<br>- Sequence Diagrams | Behavioural and Structural Modelling | - Software Architect<br>- Technology Consultant | Any software modelling tool (Microsoft Visio, Eclipse Modeling Framework, etc.) |
| | Software Architecture | Layered Architecture Model<br>- Health Sensing Layer (monitoring)<br>- Health Analytics Layer (processing) | Software Design and Architecture Model | | |
| Implementation and Validation | Patterns and Styles | Not Applicable | Not Applicable | Not Applicable | Not Applicable |
| | Tools and Technologies | - System Development<br>- System Deployment | Source Code and Service Components for Deployment | - Software Developer<br>- Software Test Engineer | - Eclipse IoT (Development)<br>- Amazon Web Services (Deployment) |
| | Validation Method | Not Applicable | Not Applicable | Not Applicable | Not Applicable |
| | Validation Context | Case Study and Scenario-based Validation (Smart Healthcare)<br>- Functional Scenarios<br>- Quality Use Cases | - Test Cases<br>- Testing Strategies | - Test Engineer<br>- Software Developer<br>- Software Architect | Eclipse IoT (Unit Testing)<br>MATLAB (Analysis and Simulation) |
| Delivery and Deployment | Deployment Type | Deployment and Configuration<br>- Front-end Health Sensor<br>- Back-end Health Analytics | Deployment Strategy | - Test Engineer<br>- Software Architect | AWS Toolkit (Amazon) |
| | Deployment Platform | Web Services (Amazon) | Not Applicable | Technology Consultant | AWS Toolkit (Amazon) |
| **Domain of Application** | | | | | |
| Smart City | The domain of application for developed systems (Health-Connect) is smart and connected health care. The system exploits on-body sensors to continuously monitor human health symptoms. The health critical data captured by sensors is processed to develop Health Profiles that can be electronically shared with other medical professionals. | | | | |

The human roles to support the software architecting activity are software architect and technology consultant. Most appropriate tool for architecting activity is Eclipse Modeling Framework (based on UML notations and profile). Tool support provides necessary automaton, while human roles enable decision support to guide and customise the process.

### C. Engineering and Development of Health-Connect System

As in Figure 5and Table 4, the initial step in system engineering and development depends on system requirements that include both the functional aspects as well as quality attributes. The main functional aspects are active real-time health monitoring, health analytics, and health recommendations that need to satisfy the quality aspects such as data security and privacy, along with system's efficiency and performance to collect, process and report the data. System modelling and design helped to model the blue-print of the system by creating an architecture model based on the Unified Modeling Language (UML) notations. Based on the platform requirements (feasibility analysis phase), UML component diagrams are used to design a layered architecture consisting of front-end sensor layer (monitoring) and back-end analytics layer (processing). After system architecting, the implementation step requires coding of the system in terms of executable modules [33]. Web-service based approach has been adopted for the development and deployment of the system. The validation phase evaluates the outlined functional and non-functional requirements as outlined earlier. System delivery and deployment exploited the distribution patterns to physically distribute the system into two layers consisting of mobile sensors and analytics server.

### D. Roles, Tools, and Technologies to Support Engineering and Development

During the engineering and development of IoT-DA systems, roles refer to human participants and their professional engagement in the system development. Human roles rely on set of tools and technologies that support automation and customisation of the SE process for system engineering and development. The case study highlights four main roles (technology consultant, software architect, software developer, and test engineer) that have been involved in the engineering and development of the system. The tool that has been used is Eclipse IDE for source code development and AWS for deploying and operationalising health care services. The tools provide semi-automation for designing and testing the system. The domain of IoT-DA system is Smart Healthcare that exploits IoT sensors to collect the health specific information. Health specific information is processed at the cloud server to generate health profile and recommendations for healthcare management.

## VI. Recommendation, Lessons Learnt, and Validity Threats

After presenting the case study, first we discuss some recommendations and lessons learnt (Section 6.1). We then discuss some validity threats as potential limitations of the research (Section 6.2)

### A. Recommendations and Lessons Learnt for Process-driven Development of IoT-DAs

Based on the analysis and experiences from the case study, we now present some recommendations and lessons learnt for process-centric SE for IoT systems in general and IoT-DA applications in particular. We consider the findings from [8, 26] for documenting the guidelines and lessons learnt based on empirical investigations in the context of software-intensive systems. Table 5 provides a structured format to present (i) recommendations as suggestive guidelines and (ii) specific lessons learnt as observations or reusable knowledge for engineering IoT-DA applications. Table 5 considers all phases of the SE development process to provide a total of 06 recommendations and 04 lessons learnt as observations. In addition to the data in Figure 5, illustrative figures (Figure 6 to Figure 9) are used to demonstrate the recommendations and lessons learnt. In the following, we exemplify one of the recommendations along with relevant observation and lessons learnt that can be referred to and used for interpreting other recommendations from Table 5. For example:

- **Recommendation:** *Perform explicit analysis of Things and Sensors during feasibility analysis.* Analysing the things and sensors can pinpoint any hardware, software, or network specific constraints in an IoT system that could impact system's functionality (e.g., data analytics) and quality (e.g., performance) attributes.

- **Example:** The Health-Connect case study (Figure 5) highlighted that health monitoring sensors (as front-end devices) lack storage and processing of computation-intensive tasks. Cloud-based servers (as back-end resources) can be used to off-load computation, storage, power-intensive tasks by exploiting mobile cloud computing technologies. However, due to small size of sensor its Wi-Fi connectivity suffers with continuous connectivity issues with frequent disconnections. This mean due to portable nature of health monitoring sensors, health sensors encounter continuous disconnection that could ultimately lead to distorted monitoring values. One of the solutions is to integrate a local gateway to the cloud (i.e., local cloudlets) that ensure better connectivity with the sensors locally as in Figure 6.

**Observation and lesson learnt:** We have illustrated our observation as Figure 6 that presents a layered architecture with front-end context-sensing sensors that is connected to back-end cloud server via a cloudlet mediator. In Figure 6, by introducing the cloudlet, sensor connectivity via Bluetooth and wireless connectivity is increased and data from sensor is managed by the local cloudlet that periodically communicates with the external cloud for storage and computation purposes. Moreover, in the Health-Connect case study, the connectivity and throughput of the health monitoring sensor is measured before and after introducing the gateway (i.e., cloudlet) between sensor and the cloud. The results are presented in Figure 7, highlighting that by integrating the cloudlets the connectivity and throughput of sensors has increased.

### B. Threats to the Validity of Research

After presenting the results of evaluation and case study, we now discuss some threats to the validity of research. The threats detailed below relate to potential limitations that need to be identified and addressed to avoid any bias and/or limitations of research findings.

**TABLE V** Recommendations and Lessons Learnt (Observations) based on Health-Connect Case Study

| Recommendations | Lessons Learnt as Observations |
|---|---|
| | **Feasibility Analysis (FA)** |

| Recommendations | Lessons Learnt as Observations |
|---|---|
| ***Objective:*** The objective of feasibility analysis phase is to analyse the operational feasibility of the system in terms of its capabilities and any constraints.<br><br>***Example:*** Sensors as embedded/portable things can provide contextual information (health, traffic, environment specific data, etc.). However, portability also introduces resource poverty (i.e., limited storage, computation, battery etc.) for context-aware sensors. Therefore, portable and embedded nature of IoT sensors, despite the offered benefits, may affect the performance and quality attributes of the system.<br><br>**Recommendation 1: Explicit Analysis for Sensors/Things of IoTs:** Case study based analysis pinpoint that health monitoring sensors (as front-end devices) lack storage and processing of computation-intensive tasks. Cloud-based servers (as back-end resources) can be used to off-load computation, storage, power-intensive tasks. However, due to limited size and capability of the of Wi-Fi sensor, there is a continuous connectivity issues with weak signals and frequent disconnections. One of the solutions is to integrate a local gateway to the cloud (i.e., local cloudlets) that ensures better connectivity with the sensors. Sensors need to occasionally communicate with the external cloud through the gateway as in Figure 6.<br><br>**Recommendation 2: Perform Hardware and Software Mapping:** The mapping is required in terms of synergising the hardware components (temperature monitoring sensor) with an associated software service (storage and analysis of sensor reading). An explicit mapping ensures that for each IoT sensor/thing there is corresponding software service to ensure storage, processing, and fault tolerance for that particular sensor. | ***Lesson Learnt – IoT system and sensors' performance increases with the integration of communication Gateway (local cloudlet):*** As in Figure 6, the connectivity and throughput of the health monitoring sensors is increased with the integration of a gateway (local cloudlet) between monitoring sensor and storage cloud. This enables the sensor to exploit Bluetooth connectivity with the local cloudlet for reliable connection. We measured the performance of the HelthConnect system and its health monitoring sensors by analysing the connectivity and throughput of the sensors that measure health conditions such as blood pressure, pulse rate, body temperature etc.<br>**-Sensors' Connectivity:** Specifically, we have measured the connectivity duration of the sensors before and after the integration of the Gateway (Figure 6).<br><br><br>**Figure 6.** Design Solution to support increased connectivity between Context Sensor (front-end) and Server (back-end)<br><br>Increased connectivity is the result of the gateway that acts as a cloudlet and enables sensors to communicate for data storage through Wi-Fi as well as Bluetooth. However, integration of the cloudlet increases connectivity overhead as the cloudlet needs to periodically connect with the external cloud to offload data to it.<br>**-Sensors' Throughput:** We also measured the sensors' throughput in terms of volume of data being produced by sensors and communicated with the gateway. The volume of data produced and communicated to the gateway is measured and logged at the gateway. Figure 7 b) shows that based on ten different trials the volume of data communicated with the gateway is larger than the data that was communicated with the cloud-server. |

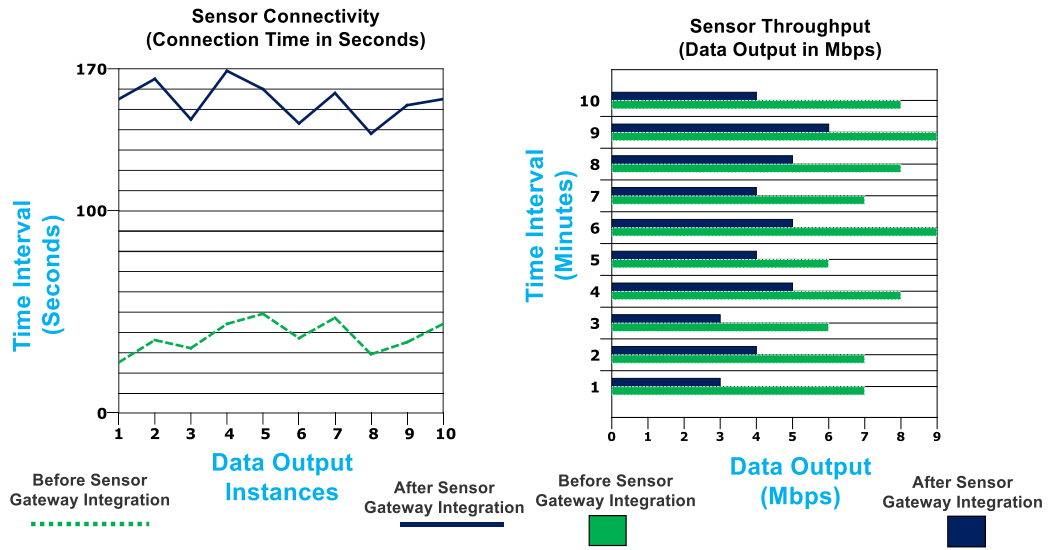

**Figure 7.** Monitoring Sensor Connectivity and Throughput after integrating of the Gateway (Local Cloudlet)

## Engineering and Development (ED)

*Objective:* To support the requirement specification, modeling, implementation, evaluation of the system to be deployed.

*Example:* The IoT-DA system needs to be engineered by specifying requirements (functional/quality attributes), creating model (design and architecture), implementing logic (source code and algorithms), and evaluating the system (validating quality requirements).

**Recommendation 3: Needs for Modeling Notations to Specify Things and Interactions:** There is a lack of tools and notations that can be used to model things and their interaction. Explicit modeling notations (UML profiles and extensions) can support the representation of Things, their interactions and distributions in an IoT-DA system as in Figure 8.

**Recommendation 4: Exploit Model-driven Development of IoT Applications:** Model-driven engineering and development needs to be adopted where model of the IoT things/sensors can be created, it needs to be mapped with software services to automatically generate the source code. Model-driven SE for IoTs can support full engineering cycle that is driven by a model that is refined at each stage from analysis, modeling, implementation, testing and deployment as highlighted in Figure 8

*Lesson Learnt* – **Modeling driven engineering enables efficiency and semi-automatio for system IoT application development:** Figure 8 demonstrates that real world context such as sensor-based human health monitoring could be represented as a model (IoT domain). A modeling notation or language helps with specification of the real world context by considering the architecturally significant requirements such as system performance, things reactivity, multi-tenancy, fault tolerance issues. By exploiting already established principle and practices of model-driven engineering, executable source code can be generated from the IoT application. Model-driven IoT relies on abstrcat models that can be incrementally refined through each phase of software development to support automation and customisation of IoT development process.

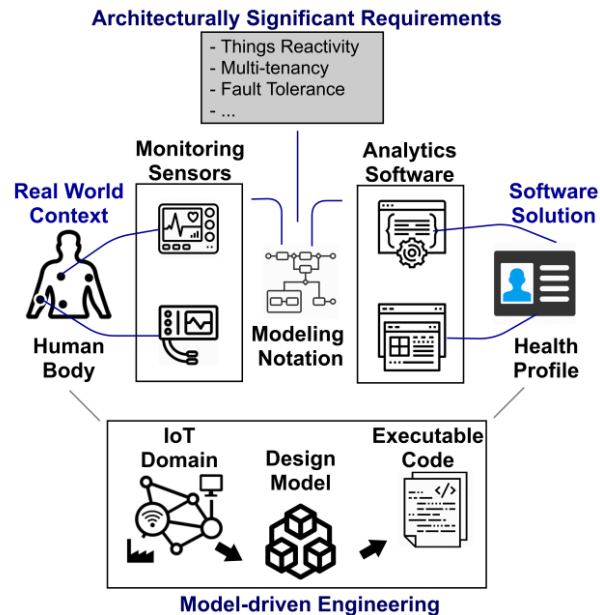

**Figure 8.** Overview of Model-driven SE for IoT-DA Application

## Roles, Tools and Technologies (RT)

*Objective:* To understand the role of tools and human decision support in the process for engineering IoT-DA applications.

**Recommendation 5: Tools Support**

*Lessons Learnt* – **Tool Support for the Process**: In order to understand the practitioners' view on the role of tool support, we asked for their feedback on 04 distinct issues related to tool support as in Figure 9. For example, the question: 'what are the primary challenges with existing tools to develop IoT systems in general and IoT-DA applications in particular?'

**Automation but Lack Process Supervision:** Tool support in the form Of IDEs, frameworks, modeling notations provides means to automate the process. Tool support eliminates the needs for laborious and effort-intensive tasks (e.g., simulation, design modeling) that can be error-prone, time consuming, and impractical. However, IoT development is an intellectual process that requires proper supervision and decision support not offered by tools that primarily focus on process automation as in Figure 9.

**Recommendation 6: Incorporate Human Decision for Process Customisation:** Process execution depends on human supervision to guide and customsie the process. Automation is ideal; however, process for engineering IoT-DA applications requires human guidance and input. Such guidance and support can be provided by various technical roles including but not limited to technology consultants, IoT developers, and IoT test engineering as highlighted in Figure 9. In particular, the role of technology consultants is critical to guide the development process (as an integration engineer) between hardware and software development teams.

**Recommendation 7: Practitioners' View on Tool Support and Human Roles in the Process**
In order to objectively understand the role of tool support and human decision in the SE process, we engaged a total of 08 practitioners from Health-Connect case study to seek their feedback. The practitioners were asked for their input in a questionnaire based on 07 points (04 on tool support and 03 on human roles) as illustrated in Figure 9. The horizontal axis represents the percentage of response, whereas the vertical axis present questions being asked from the practitioners of the case study. The responses as practitioners' view are presented in Figure 9 that highlights the needs, challenges, and role of tool-based automation and human-driven decision support for SE processes in the context of IoT-DA applications.

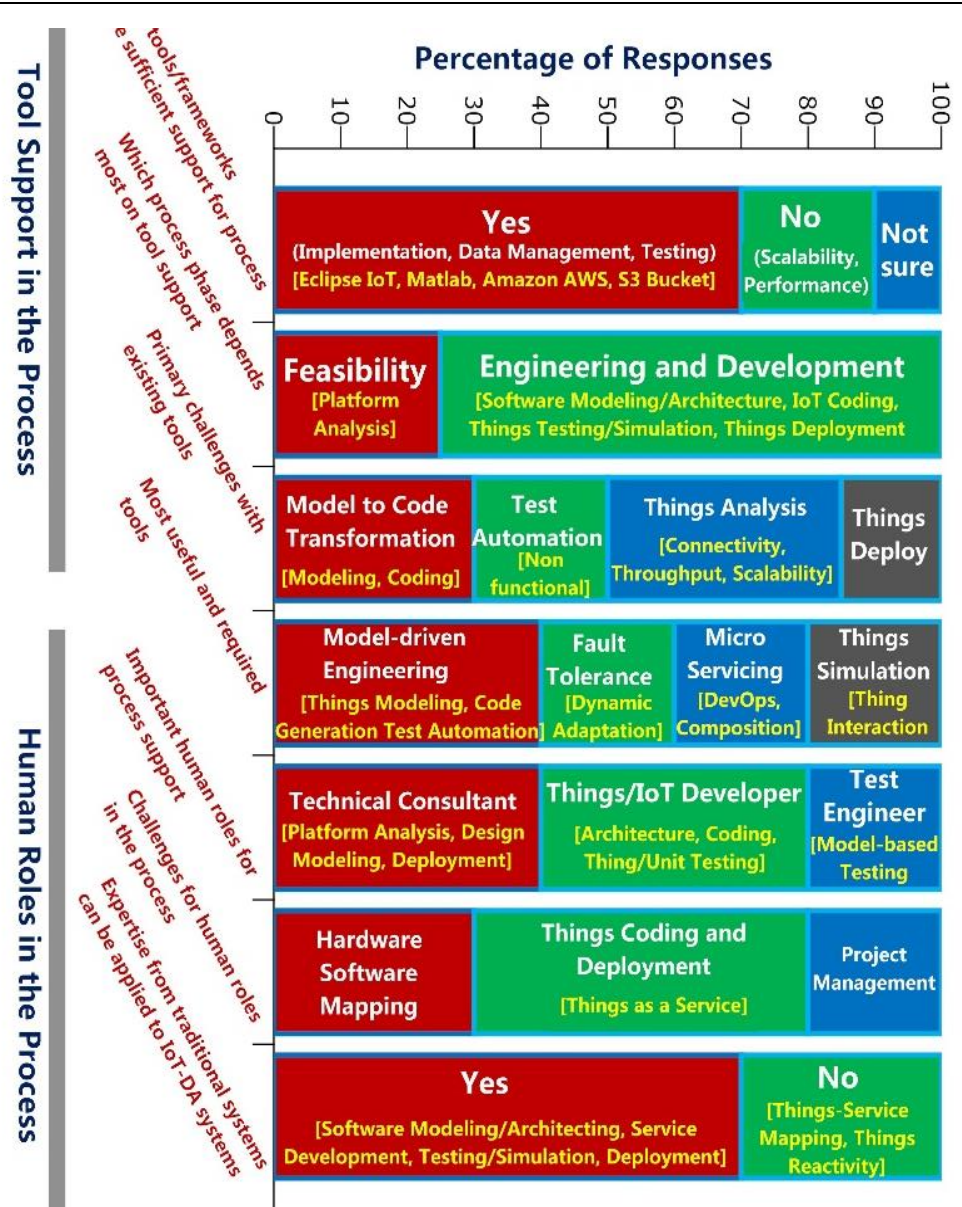

**Figure 9.** Overview of Practitioners' Responses on Tool and Human Support

The answers highlighted primary challenges that include but not limited to following four:
- *Model to Code Transformation* that requires creating a domain model that could be transformed into executable source code with mode-driven development
- *Test Automation* to validate non-functional (a.k.a. architecturally significant requirements) such as performance, scalability, connectivity, etc.
- *Things Analysis* by means of simulation or thing's interaction monitoring to understand the connectivity, throughput, scalability of things and sensors
- *Things Deployment* and their configurations for a functional system.

*Lessons Learnt* – **Human Decision in the Process**: The role of human decision and intervention is being analysed based on 03 distinct questions as in Figure 9. For example, corresponding to the question: 'what are the most important roles in SE process for IoT-DA applications?', the feedback from practitioners' highlighted:
- *Technical Consultant:* to support activities such as platform analysis, design modeling, IoT deployment, etc.
- *Things/IoT Developer:* is mainly concerned about implementing IoT applications based on architecting, coding and unit-testing the things.
- *Test Engineer:* to perform validation of the system in the context of functional and non-functional requirements

**Threat I - Identification of Process-centric Information**
We have used systematic mapping study process to identify the relevant processes based on published academic research and publicly available industry scale solutions. Systematic mapping study is based on evidence-based software engineering process that provides a systematic process to identify and qualitatively select the literature/domain under investigation (Figure 3). Based on the guidelines of mapping study, we tried our best to identify the most relevant processes based on multistep process (Section 3, Figure 3). However, some of the non-publicly available processes were not identified and ultimately not included in the evaluation. In this context, future work can focus on accommodating more processes in the process repository to avoid the potential bias in process identification.

**Threat II - Documenting and Presenting the Results**
After identification, we applied thematic process (Section 3) to document the process information using evaluation framework (Figure 3) and process matrix (Figure 4, Table 3). Evaluation framework and process matrix have been derived based on some well-defined taxonomies and evaluation framework from software engineering. After developing the framework, we sought domain experts' feedback to refine and finalise the framework and process matrix. However, there is a need to extend the criteria for documenting and evaluating the processes that could be undertaken as part of future research.

**Threat III – Limited Data with Single Case Study**
We analysed the evaluation framework and process activities by presenting smart healthcare case study on process-centric development of IoT-DA applications. Case study provides practical scenarios and use cases to demonstrate process and its underlying activities. The case study has been selected from an IoT solution provider with a team of 07 people. The case study represents an industrial system, however; single case study provides limited data for more diverse and fine-grained analysis. Future research requires more case studies to be investigated. Also, a survey involving different practitioners (geographically distributed teams) can be conducted to seek practitioners' feedback on SE processes and practices for IoT-DA applications.

## VII. Conclusions
IoTs represent a class of software-intensive systems that exploit sensors and networking technologies to enable data-driven intelligence for smart cities and infrastructures. Data analytics systems and applications that ingest data from IoTs exploits sensors as interconnected things to collect, exchange, and process contextual data - supporting data-driven intelligence and enabling decision support - for smart systems. IoT-DA applications that exploit IoT platforms and technologies rely on an engineering life-cycle that supports systematic design, development, deployment and evolution of such applications. Software engineering for IoT-DA unifies science and engineering of analysing data (using software theories and algorithms) and transforming raw data into useful information by (using software tools and technologies). SE for IoT-DA aims to strengthen strategic capabilities of enterprises by exploiting software tools, algorithms, and applications that collect data from IoT sensors and systems and deliver intelligent analytics to stakeholders. However, a complex blend of hardware and software artefacts poses challenges for engineering and development of dynamic IoT applications that require reusable knowledge and best practices of software and system development.

The proposed research complements community-wide initiatives on exploiting engineering practices, processes, patterns, frameworks, and tools for developing IoT-driven systems and software. We have used evidence-based software engineering approach to develop a criteria-based framework that streamlines and objectively evaluates software engineering processes and practices for IoT-DA applications. This article empirically derives an evaluation framework to systematically investigate the role of SE processes and their underlying practices for engineering IoT-DA applications. We applied systematic mapping process to identify a total of 37 processes with qualitative selection and investigation of 16 processes that support SE for IoT-DA based on academic research and industrial solutions. Results of evaluation and case study based validation highlight open challenges, role of process automation, human decisions, lessons learnt, and recommended practices for engineering and development of IoT-DA applications. Primary contributions of the proposed research include:

- Establishing a process-centric view in terms of SE research and practices that exploit models, patterns, framework and tool support to design, develop, deploy, and maintain software-intensive systems and applications of IoT driven data analytics.

- Empirically developed framework for criteria-based evaluation of strengths, limitations, and open challenges pertaining to research and development of IoT driven data analytics. Case study based approach is used to validate the framework in a practical context.

Creating a repository of SE processes – structured catalogue of solutions and development activities – that encapsulate best practices.

## Appendix A

List of Processes (Academic Publications and Industrial Solutions) to Support SE for IoT-DA

| Process ID | Process Name/Title | Year of Publication/ Availability | Process Source |
|---|---|---|---|
| P01 | CityPulse: Large Scale Data Analytics Framework for Smart Cities | 2016 | Academic Publication |
| P02 | A Big Data Analytics Architecture for the Internet of Small Things | 2018 | Academic Publication (Industry-Academia Collaboration) |
| P03 | CLOTHO: A Large-Scale Internet of Things based Crowd Evacuation Planning System for Disaster Management | 2018 | Academic Publication |
| P04 | Cloud-assisted Industrial Internet of Things (IIoT)-enabled Framework for Health Monitoring | 2016 | Academic Publication |
| P05 | Urban Planning and Building Smart Cities based on the Internet of Things using Big Data Analytics | 2016 | Academic Publication |
| P06 | Analytics Everywhere: Generating Insights From the Internet of Things | 2019 | Academic Publication |
| P07 | Designing a Smart Transportation System: An Internet of Things and Big Data Approach | 2019 | Academic Publication |
| P08 | Internet of Things Mobile-Air Pollution Monitoring System (IoT-Mobair) | 2018 | Academic Publication |
| P09 | TCS Sensor Data Analytics IoT Framework | 2017 | Industrial Solution (TATA Consultancy Services) |
| P10 | IoT Analytics - ThingSpeak Internet of Things | 2010 | Industrial Solution (ThingSpeak) |
| P11 | The Industrial Internet of Things Industrial Analytics Framework | 2017 | Industrial Solution (Industrial Analytics Consortium) |
| P12 | SeeboIoT One | - | Industrial Solution (Seebo) |
| P13 | Fujitsu Cloud IoT Platform | 2016 | Industrial Solution (Fujitsu) |
| P14 | AWS IoT Analytics | 2018 | Industrial Solution (Amazon) |
| P15 | Google Cloud IoT | 2008 | Industrial Solution (Google) |
| P16 | Azure IoT | 2015 | Industrial Solution (Microsoft) |